\documentclass[11pt,a4paper]{article}
\usepackage[top=0.85in,left=1.in,right=1.in,footskip=0.75in]{geometry}

\usepackage{amsmath,amssymb}

\usepackage{textcomp,marvosym}

\usepackage{cite}

\usepackage{nameref,hyperref}

\usepackage[nopatch=eqnum]{microtype}
\DisableLigatures[f]{encoding = *, family = * }

\usepackage[table]{xcolor}

\usepackage{bm}

\usepackage{float}
\usepackage{amsmath}
\usepackage{booktabs}

\usepackage{array}

\newcolumntype{+}{!{\vrule width 2pt}}

\usepackage[aboveskip=1pt,labelfont=bf,labelsep=period,justification=raggedright,singlelinecheck=off]{caption}

\makeatletter
\renewcommand{\@biblabel}[1]{\quad#1.}
\makeatother

\usepackage{lastpage,fancyhdr,graphicx}
\usepackage{epstopdf}

\begin{document}
\vspace*{0.2in}

\begin{flushleft}
{\Large
\textbf\newline{Speaker-Independent Dysarthria Severity Classification using Self-Supervised Transformers and Multi-Task Learning} 
}




Lauren Stumpf\textsuperscript{1*},
Balasundaram Kadirvelu\textsuperscript{1*},
Sigourney Waibel\textsuperscript{1},
A Aldo Faisal\textsuperscript{1,2§},
\\
\bigskip
\textbf{1} Brain \& Behaviour Lab, Department of Computing and Department of Bioengineering, Imperial College London, London, United Kingdom
\\
\textbf{2} Chair in Digital Health, Faculty of Life Sciences, University of Bayreuth, Bayreuth, Germany
\\
\bigskip

%
%

\textsuperscript{*} Authors contributed equally to this work \newline
\textsuperscript{§} a.faisal@imperial.ac.uk

\end{flushleft}

\section*{Abstract}

Dysarthria, a condition resulting from impaired control of the speech muscles due to neurological disorders, significantly impacts the communication and quality of life of patients. The condition's complexity, human scoring and varied presentations make its assessment and management challenging. This study presents a transformer-based framework for automatically assessing dysarthria severity from raw speech data. It can offer an objective, repeatable, accessible, standardised and cost-effective and compared to traditional methods requiring human expert assessors. We develop a transformer framework, called Speaker-Agnostic Latent Regularisation (SALR), incorporating a multi-task learning objective and contrastive learning for speaker-independent multi-class dysarthria severity classification. The multi-task framework is designed to reduce reliance on speaker-specific characteristics and address the intrinsic intra-class variability of dysarthric speech. We evaluated on the Universal Access Speech dataset using leave-one-speaker-out cross-validation, our model demonstrated superior performance over traditional machine learning approaches, with an accuracy of $70.48$\% and an F1 score of  $59.23\%$. Our SALR model also exceeded the previous benchmark for AI-based classification by $16.58\%$. We open the black box of our model by visualising the latent space where we can observe how the model substantially reduces speaker-specific cues and amplifies task-specific ones, thereby showing its robustness. In conclusion, SALR  establishes a new benchmark in speaker-independent multi-class dysarthria severity classification using generative AI. The potential implications of our findings for broader clinical applications in automated dysarthria severity assessments.

\section*{Author summary}
Traditional methods for evaluating dysarthria—a speech impairment caused by neurological disorders like Parkinson's disease and cerebral palsy—are often subjective, relying heavily on expert opinions. This highlights the need for more objective and standardised assessment tools. To meet this need, we have developed a new deep learning framework that uses wav2vec 2.0, a speech model originally trained on speech from healthy individuals, to classify levels of dysarthria severity. Our approach combines a contrastive loss task with classifying the severity of dysarthria. Our results demonstrate that this framework is more effective than traditional methods, setting a new standard for dysarthria severity classification. This has significant potential to enhance dysarthria assessments in clinical settings.


\section*{Introduction}
Dysarthria, characterised by impaired control over speech muscles due to neurological conditions, has a profound impact on communication and quality of life \cite{darley1969differential}. Various neurological disorders, including cerebral palsy, Parkinson's disease, and amyotrophic lateral sclerosis (ALS), manifest dysarthria, leading to a spectrum of speech abnormalities \cite{whitehill2000speech, scott1983speech}. The complex nature of dysarthria, influenced by underlying pathology and individual patient characteristics, presents significant challenges in both assessment and management \cite{joy2018improving}. Effective assessment of dysarthria is crucial not only for understanding its severity but also for monitoring disease progression and tailoring therapeutic interventions \cite{freed2018motor}.

The traditional approach to dysarthria assessment involves auditory-perceptual evaluations by experienced speech-language pathologists. However, this method is subjective and may lack consistency, underlining the need for more objective and standardised assessment tools \cite{gavidia1996direct}. With advancements in technology, automated, machine learning-based tools have emerged as promising alternatives, offering the potential for more objective, efficient and accessible dysarthria assessments which can be especially advantageous for individuals facing mobility challenges due to co-occurring physical disabilities \cite{baghai2012automatic}. 

The development of accurate automated assessment tools for dysarthria is a complex task. The variability in speech patterns among individuals with dysarthria, influenced by the underlying neurological condition and severity of the speech disorder, presents a significant challenge. Furthermore, the lack of extensive dysarthric speech datasets, exacerbated by the difficulties in collecting prolonged speech samples from individuals with severe dysarthria, hampers the training of advanced machine learning models that require large amounts of data \cite{hawley2007speech, christensen2012comparative}.

Recent advancements in deep learning, particularly transformer models, have shown potential in various speech processing tasks \cite{vaswani2017attention, gong2021ast}. Their ability to capture contextual information across entire input sequences makes them well-suited for modelling the nuanced effects of dysarthria on speech. However, applying these models, primarily developed for healthy speech tasks, to speech signals of those affected by dysarthria, is still a challenging endeavour.

In this study, we propose a novel framework that leverages a transformer model trained on healthy speech to assess the severity of dysarthria. Our methodology exploits the wav2vec 2.0 \cite{baevski2020wav2vec} model, a state-of-the-art self-supervised transformer model, to extract meaningful speech representations. Through self-supervised pre-training on healthy speech, the wav2vec 2.0 model acquires an understanding of speech's fundamental structure, a characteristic we leverage in our framework to overcome data scarcity constraints. Furthermore, our framework incorporates a multi-task learning strategy focusing on phoneme prediction to prevent over-fitting and to accommodate the inherent intra-class variability observed in dysarthric speech. 
Through rigorous evaluation and validation, we demonstrate the effectiveness of our proposed framework, thereby contributing to the advancement of more accurate and accessible assessments for dysarthria.

\section*{Materials and Methods}

\subsection*{Dataset}
We used the Universal Access dysarthric speech corpus (UA-Speech) \cite{kim2008dysarthric}, a comprehensive and commonly used English language dataset for dysarthric speech research. Unlike many other datasets, UA-Speech provides both word-level and phoneme-level transcriptions, essential for our research methodology. The dataset includes recordings from 15 subjects with dysarthria and 13 age-matched healthy controls of the following spoken words: 10 digits, 26 letters, 19 computer commands, 100 common words (repeated three times) from the Brown Corpus, and 300 uncommon words from Project Gutenberg novels chosen to maximise biphone diversity. These recordings were captured using a seven-channel microphone array and five native American English speakers transcribed the recordings. Each subject's speech intelligibility was calculated based on the average percentage of words correctly transcribed. Subjects were categorised into four levels of dysarthric severity based on their intelligibility ratings: very low severity(76-100\% intelligible), low severity (51-75\% intelligible), medium severity (26-50\% intelligible), and high severity (0-25\% intelligible). For a detailed overview of UA-Speech, readers are referred to \cite{kim2008dysarthric}.

\subsection*{Finetuning the wav2vec model} 
We chose to use wav2vec 2.0 \cite{baevski2020wav2vec} for our pretrained transformer over other models like Audio Spectrogram Transformer \cite{gong2021ast} and HuBERT \cite{hsu2021hubert} based on empirical evidence from early experimentation. The wav2vec 2.0 model was pretrained on an expansive 960-hour dataset from diverse audio-book libraries and the entire pretraining process was distributed across 64 V100 GPUs and spanned 1.6 days. The underlying transformer architecture of this model consists of 12 transformer blocks. Each block has a model dimension of 768, an inner feed-forward network dimension of 3072, and 8 attention heads.  We utilise the \texttt{facebook/wav2vec2-base} model available from the 4.33.1 version of the HuggingFace library \cite{wolf2019huggingface}. To fine-tune our model for the specialised task of classifying dysarthria severity, we added a linear classification head comprising two linear layers with a ReLU activation. The fine-tuning training was conducted with a batch size of four, using the Adam optimiser. The optimiser was set with a learning rate of \(0.0005\), betas configured to \( (0.9, 0.98) \), and an epsilon value of \(1 \times 10^{-8}\).

\subsection* {Speaker-Agnostic Latent Regularisation (SALR) Framework} \label{subsec:SALR}

Our initial experiments demonstrated a significant challenge with simply fine-tuning an off-the-shelf wav2vec2.0 model: its tendency to overfit to specific speakers. This could be attributed to the limited diversity within the UA-Speech dataset, which only includes 15 distinct dysarthric speakers. Instead of effectively learning the characteristics specific to dysarthria severity, the model appears to be leveraging speaker-specific cues. This approach can minimise the training loss, through recognising the speaker’s identity and subsequently assigning a dysarthria severity label. But this approach struggles with new, previously unheard speakers, highlighting a gap in the model's ability to generalise. This issue extends to the latent space, potentially leading to the formation of speaker-centric clusters. Different words uttered by the same speaker are more closely embedded in the latent space compared to the same words spoken by different speakers, even if those speakers have the same level of dysarthria severity. This entangled representation of words is because the complexity of a word — defined by its syllables, phonetic structures, and the necessary motor control for pronunciation — directly impacts how prominently dysarthric symptoms manifest. Without a clear representation in the latent space that accounts for word complexity, the model faces challenges.

To address these issues in the latent space, we introduce a regularisation contrastive loss framework, which we call Speaker-Agnostic Latent Regularisation (SALR), to disentangle the embeddings. Our framework represents a specialised configuration that enhances the base wav2vec 2.0 model with additional components tailored to accomplish an auxiliary task alongside the primary dysarthria classification.
The auxiliary task in this framework is a contrastive learning task which aims to ensure that the separation between word embeddings within a shared severity classification becomes speaker-agnostic, thereby helping prevent the model from learning embeddings that embed speaker-specific characteristics. Specifically, the framework consists of an extra head designed for the auxiliary task, a weighted loss function crafted to balance the learning objectives of both the primary and auxiliary tasks, and a training regimen that specifies how the weighted loss function is applied.

To illustrate this framework, let \(E_{AX}\) represent the embedding of a `Word A' articulated by `Patient X', \(E_{BX}\) for `Word B' spoken by `Patient X', and \(E_{BY}\) for `Word B' spoken by `Patient Y', where both `Patient X' and `Patient Y' share the same dysarthria severity label. We define the distances 
$d_1: \text{Distance between } E_{AX} \text{ and } E_{BX}$ and
$d_2: \text{Distance between } E_{AX} \text{ and } E_{BY}.$

The objective is to make \(d_1\) approximately equal to \(d_2\). This ensures that within a particular severity classification, variations in embeddings stem from the words themselves, not the speakers. Initially, we anticipate that \(d_1\) will be smaller than  \(d_2\). This is because the distance \(d_1\) captures the distance between words spoken by the same speaker, and as previously discussed, our latent space tends to be influenced by speaker-specific traits. 

To achieve our objective, we utilise triplet margin loss \cite{dong2018triplet} with specific aims. First, we intend to push away \(E_{BX}\) from the anchor \(E_{AX}\) by considering \(E_{BX}\) as the negative sample and \(E_{AX}\) as the anchor. Given that these embeddings originate from the same speaker, we expect them to be closely located in the latent space. Thus to balance \(d_1\) and \(d_2\), the distance \(d_1\) needs to be expanded. Simultaneously, we aim to pull \(E_{BY}\) closer to \(E_{AX}\) by designating \(E_{BY}\) as the positive sample and retaining \(E_{AX}\) as the anchor. Since these embeddings are from different speakers, we hypothesise that their distance in the latent space will be larger. Thus to equate \(d_1\) and \(d_2\), the distance \(d_2\) should be contracted. We note that the triplet margin loss is designed to ensure the anchor embedding, \(E_{AX}\), is nearer to the positive sample \(E_{BY}\) than to the negative sample \(E_{BX}\), by a specific distance known as the margin, \(m\). However, by intentionally keeping \(m\) minimal and taking into account the initial distances between the embeddings, we aim to make the distances between the anchor-positive and anchor-negative pairs approximately the same and get rid of the initial disparity. 

We hypothesise that implementing this regularising loss will be beneficial because it shifts the model's focus from identifying speakers to distinguishing words. Specifically, the model should be able to differentiate between two words regardless of whether they are spoken by the same person or by different individuals with the same dysarthria severity. For example, by creating a greater distance between the anchor embedding \(E_{AX}\) and \(E_{BX}\), the model is forced to learn an embedding that focuses on the differences between the two words rather than relying on the speaker's identity thereby making the embeddings speaker-independent given a severity class. This enhanced ability to discriminate between words allows for more accurate comparisons as it can disentangle the complexity of the word from the dysarthria.

\begin{figure}[H]
    \centering
    \begin{minipage}{0.9\textwidth}
  \includegraphics[width=1.0\textwidth]{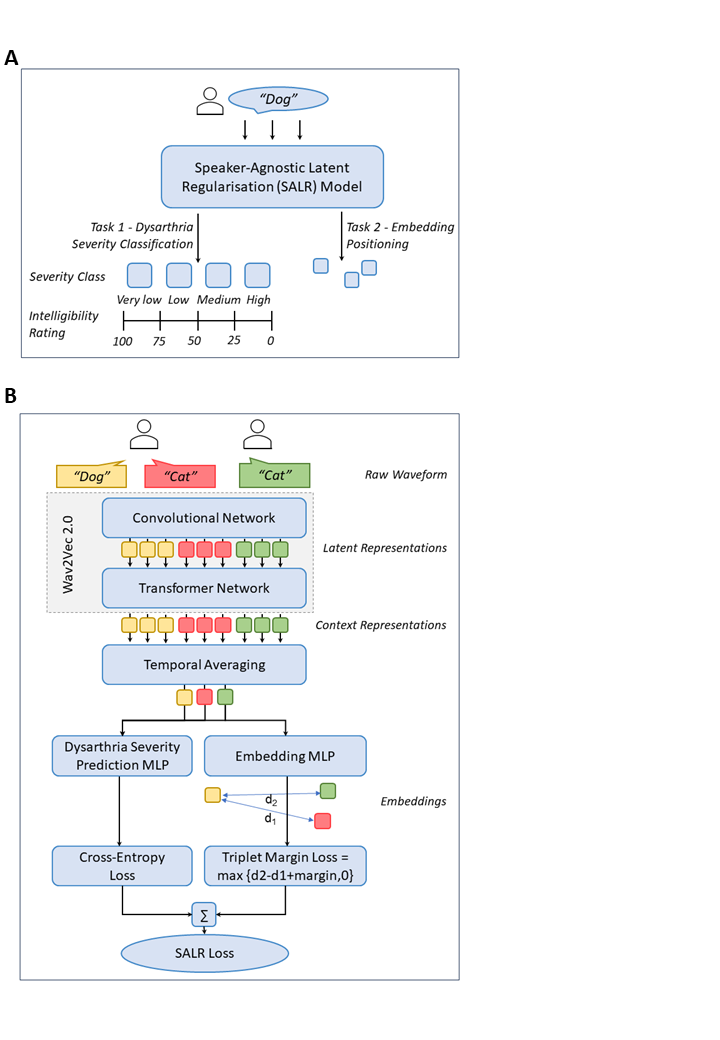}
    \caption{Architecture \textbf{A.} Overview of the Speaker-Agnostic Latent Regularisation (SALR) Framework. \textbf{B.} Detailed Visualisation of the Architecture and Training Procedure of the Speaker-Agnostic Latent Regularisation (SALR) Framework. }
    \label{fig:methods-arch}
    \end{minipage}
\end{figure}

Fig~\ref{fig:methods-arch} provides an architectural overview of the SALR framework. The implementation adds an additional embedding head to the base model. Training configurations mirror those of the base model, with modifications to the loss function: a weighted combination of Triplet Margin Loss and Cross Entropy Loss.

The triplet margin loss ($\text{TML}$) is defined as:
\begin{equation}
    \text{TML}(E_{AX}, E_{BY}, E_{BX}) = \max(0, \quad d(E_{AX}, E_{BX}) - d(E_{AX}, E_{BY}) + m)
    \label{eq:triplet_margin_loss}
\end{equation}

In Eq.\ref{eq:triplet_margin_loss}, \(E_{AX}\) acts as the anchor, \(E_{BY}\) is the positive sample, and \(E_{BX}\) is the negative sample. The term \(m\) is a predefined margin set to 0.05. The distance function \( d(x, y) \) is chosen to be the \(L_2\) Euclidean distance.

The final loss \( L \) is expressed as 
$
L = \epsilon L_{\text{reg}} + \gamma L_{\text{CE}}
$, where \( \epsilon \) serves as the weighting parameter, and is set at 0.01. The parameter \( \gamma \) starts at 0 for 3000 steps, allowing the model to focus on contrastive regularisation. It is then updated to 1, incorporating cross-entropy loss into the training regimen. These parameters were determined through experimentation on the 10 control patients, data we do not use in training or testing. This helps to show the robustness and generalisability of our approach, as the parameters were not tuned on data from the control group, thereby validating its potential for real-world clinical applications.

\subsection* {Baseline models} 

To contextualise our findings, we compare our results with three established baselines in dysarthria severity classification, namely: XGBoost~\cite{chen2016xgboost}, Multi-layer Perceptron (MLP)~\cite{murtagh1991multilayer}, and CNN-LSTM~\cite{shih2022dysarthria}. XGBoost is a gradient-boosted decision tree algorithm designed for speed and performance, which excels in classification and regression tasks. It iteratively corrects the mistakes of the previous trees, and the final prediction is the sum of the predictions from all the trees. Multi-layer Perceptron (MLP) is a class of feedforward artificial neural networks, consisting of at least two layers of nodes. Each node is a neuron with a nonlinear activation function. CNN-LSTM is a hybrid neural network model that combines Convolutional Neural Networks (CNNs) and Long Short-Term Memory networks (LSTMs). This model is capable of capturing both spatial and temporal dependencies in data, making it particularly suitable for sequence prediction problems.

For models requiring tabular data (XGBoost and MLP), we utilised the extended Geneva Minimalistic Acoustic Parameter Set (eGeMAPS), an extensive feature set of 88 acoustic features for speech analysis. For the LSTM-CNN model, we opted for an end-to-end learning approach with spectrograms as input. Spectrograms are visual representations of the spectrum of frequencies of a signal as they vary with time, and are especially useful for capturing the irregularities in dysarthric speech.

To optimise the hyperparameters of our baseline models, we employed Bayesian optimisation using the Optuna library~\cite{akiba2019optuna}. This method systematically explores the hyperparameter space to find the optimal set, balancing exploration and exploitation. We conducted 100 trials for each model using the 10 control patients to fine-tune the hyperparameters - once again reserving the dysarthric data for training and testing.


\section*{Results}

\subsection*{Speaker-Dependent vs Speaker-Independent Splits}
In our investigation, we examined both speaker-dependent and speaker-independent data splits for training and evaluating models for dysarthric speech severity classification. The speaker-dependent split facilitates the model's training and testing on data from the same individuals, albeit with varying words during the testing phase. Although this method aids in model training due to the consistency of voice patterns, its applicability in a clinical environment is restricted, as it fails to validate the model against new patients—a fundamental requirement for an automated diagnostic tool. Consequently, our emphasis was placed on the speaker-independent data split setup, in which the model is trained and assessed on data from distinct groups of speakers. This ensures the model's capacity to generalise across unfamiliar voices. Our study presents findings on the speaker-independent multi-class severity classification task, requiring the model to categorise the severity of dysarthric speech into four distinct levels: 'Very Low', 'Low', 'Mid', and 'High'. This approach is vital as it aligns directly with clinical relevance and the model’s efficacy across diverse patient conditions.

\subsection*{Speaker-Independent Multi-Class Severity Results}
We implemented a leave-one-speaker-out cross-validation scheme, utilising 465 utterances from 14 speakers for training and 300 distinct words from a single speaker for testing, thus ensuring the exclusion of the test speaker's data from the training set. This methodology was systematically repeated across five iterations, with both the mean and standard deviations recorded, as delineated in Tables \ref{tab:multi_class_speaker_independent} and \ref{tab:multi_class_speaker_independent_f1}. This rigorous approach underscores the robustness and reliability of our research findings.

Tables \ref{tab:multi_class_speaker_independent} and \ref{tab:multi_class_speaker_independent_f1} delineate the performance metrics of our proposed frameworks in comparison with established baseline models, such as XGBoost, MLP, and LSTM-CNN. The baseline models demonstrated suboptimal performance, each yielding an accuracy below 50\%. Conversely, the finetuned wav2vec base model exhibited an accuracy of 64.81\%. Notably, our innovative SALR framework surpassed all comparative models, including the finetuned wav2vec 2.0 model, achieving the highest accuracy of \(70.48 \pm 1.11\), a finding corroborated by the F1 scores presented in Table \ref{tab:multi_class_speaker_independent_f1}.

Despite the good overall performance of the framework models, a disparity between accuracy and F1 scores was observed, indicating inconsistent classification performance across different severity classes. Despite the frameworks' favourable performance, a noted discrepancy between accuracy and F1 scores highlighted the inconsistency in classification across different dysarthric severity levels. While aggregate metrics such as F1 score and accuracy offer substantial insights, further granularity is achieved through the analysis of confusion matrices.

\begin{figure}[H]
    \includegraphics[width=0.9\textwidth]{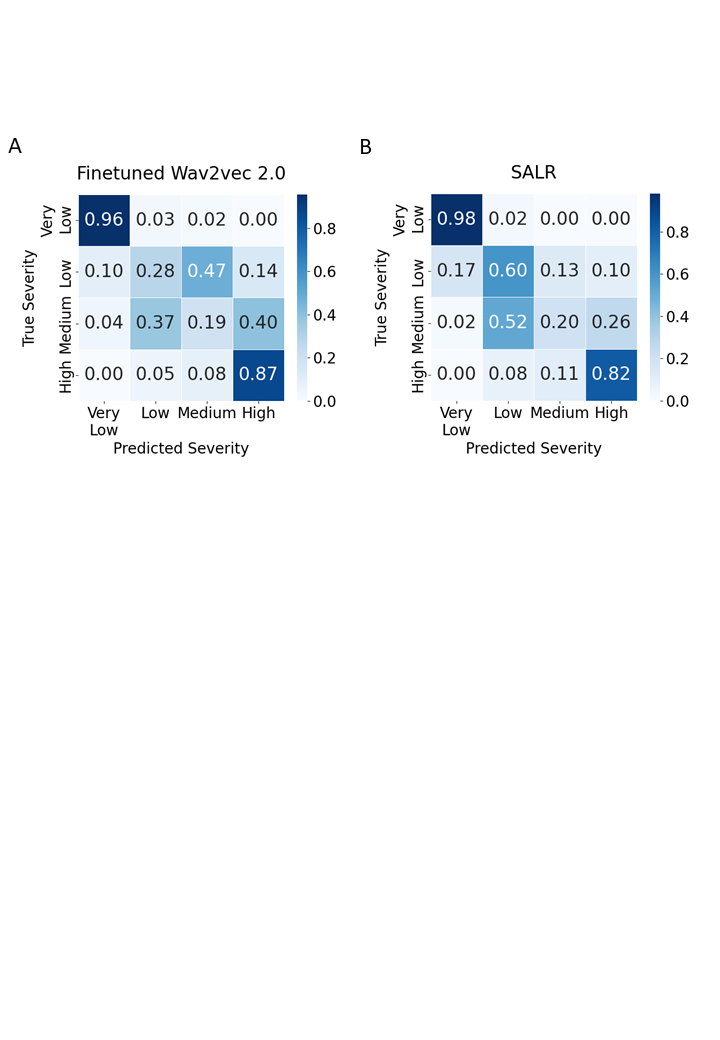}
    \caption{Normalised confusion matrices of \textbf{A.} fine-tuned Wav2Vec 2.0 model, \textbf{B.} SALR framework}
    \label{fig:cm}
\end{figure}

The confusion matrices (Fig. \ref{fig:cm}) illustrated the models' proficiency in classifying extreme dysarthric severities; however, they also highlighted significant challenges in differentiating between 'Low' and 'Mid' severity classes. Specifically, the fine-tuned wav2vec 2.0 model (Fig. \ref{fig:cm}A) frequently misclassified instances from 'Low' as 'Mid' severity, and vice versa for 'Mid' classified as 'High'. Analogously, the SALR framework (Fig. \ref{fig:cm}B) encountered difficulties in accurately distinguishing 'Low' from 'Mid' severity cases. These challenges are primarily attributable to the limited data available for these categories post-segmentation, with only two patients remaining in each of the 'Low' and 'Mid' categories, thereby limiting the models' learning efficacy. Compounding this issue is the lack of distinct boundaries between these classes. For instance, patient M16, categorised under 'Mid' severity with an intelligibility rating of 43\%, stands in stark contrast to other individuals within the same category, such as M07 and F02, who have ratings of 28\% and 29\%, respectively. Conversely, the lowest-rated individual in the adjacent 'Low' category, M05, had a rating of 58\%. This disparity in intelligibility ratings approximately equates the gap between M16 and either its own category or the neighbouring 'Low' category, thereby blurring the classification boundaries. Consequently, the models' ability to accurately classify ambiguous cases like M16 may be compromised due to the combined factors of data sparsity and ambiguous class distinctions.

\begin{table}[htbp]
    \centering
    \caption{Table presenting mean accuracy scores for each patient across five runs of leave-one-out speaker evaluation}
    \label{tab:multi_class_speaker_independent}

    \begin{tabular}{cccccc}
    \toprule
    Patient Code & MLP & LSTM-CNN & XGBoost & \begin{tabular}[c]{@{}c@{}}Finetuned \\ wav2vec 2.0\end{tabular} & SALR\\
    \midrule
    M04 & 51.78 $\pm$ 1.84 & 73.49 $\pm$ 2.34 & 62.83 $\pm$ 2.98 & 87.19 $\pm$ 0.95 & 79.62 $\pm$ 0.79 \\
    F03 & 57.39 $\pm$ 3.22 & 70.32 $\pm$ 4.33 & 61.91 $\pm$ 2.43 & 89.46 $\pm$ 0.84 & 77.00 $\pm$ 0.67 \\
    M12 & 61.89 $\pm$ 1.84 & 74.47 $\pm$ 2.33 & 71.89 $\pm$ 2.42 & 92.40 $\pm$ 0.74 & 88.64 $\pm$ 0.93 \\
    M01 & 56.74 $\pm$ 2.33 & 60.83 $\pm$ 3.42 & 63.84 $\pm$ 1.89 & 78.13 $\pm$ 1.09 & 81.18 $\pm$ 0.93\\
    M07 & 15.98 $\pm$ 5.84 & 7.38 $\pm$ 4.84 & 10.18 $\pm$ 4.33 & 7.67 $\pm$ 1.89 & 20.00 $\pm$ 1.57\\
    F02 & 9.39 $\pm$ 3.84 & 5.38 $\pm$ 3.33 & 7.85 $\pm$ 5.75 & 22.74 $\pm$ 1.89 & 21.20 $\pm$ 2.39 \\
    M16 & 13.73 $\pm$ 4.39 & 8.48 $\pm$ 4.72 & 11.43 $\pm$ 3.27 & 26.02 $\pm$ 2.00 & 19.12 $\pm$ 1.32 \\
    M11 & 8.48 $\pm$ 3.28 & 13.49 $\pm$ 5.47 & 13.43 $\pm$ 4.33 & 32.49 $\pm$ 1.04 & 58.10 $\pm$ 1.32 \\
    F04 & 6.48 $\pm$ 3.82 & 7.43 $\pm$ 4.37 & 11.85 $\pm$ 4.46 & 25.83 $\pm$ 2.38 & 61.60 $\pm$ 1.01 \\
    M05 & 9.04 $\pm$ 4.84 & 9.49 $\pm$ 5.79 & 16.89 $\pm$ 4.23 & 26.58 $\pm$ 1.00  & 60.53 $\pm$ 1.39 \\
    M09 & 73.38 $\pm$ 2.38 & 72.78 $\pm$ 2.80 & 81.89 $\pm$ 2.89 & 92.58 $\pm$ 0.89 & 98.43 $\pm$ 0.89 \\
    M08 & 72.37 $\pm$ 2.80 & 80.41 $\pm$ 3.24 & 78.94 $\pm$ 1.98 & 95.55 $\pm$ 0.79 & 97.20 $\pm$ 0.71 \\
    M10 & 76.47 $\pm$ 1.98 & 75.49 $\pm$ 1.32 & 81.89 $\pm$ 2.89 & 96.05 $\pm$ 1.84 & 97.70 $\pm$ 0.90  \\
    M14 & 77.90 $\pm$ 2.81 & 79.95 $\pm$ 1.39 & 79.80 $\pm$ 1.39 & 97.98 $\pm$ 0.67 & 98.10 $\pm$ 0.90 \\
    F05 & 74.24 $\pm$ 3.81 & 72.38 $\pm$ 1.43 & 80.84 $\pm$ 1.23 & 95.36 $\pm$ 0.84 & 98.80 $\pm$ 0.93\\
    \midrule
    Average & 44.35 $\pm$ 3.27 & 47.45 $\pm$ 3.34 & 49.03 $\pm$ 2.98 & 64.81 $\pm$ 1.26 & \textbf{70.48} $\pm$ 1.11  \\
    \bottomrule
    
\end{tabular}
\end{table}

\begin{table}[htbp]
    \centering
    \caption{Table presenting mean F1 score across five runs of leave-one-out speaker evaluation}
    \label{tab:multi_class_speaker_independent_f1}
    \begin{tabular}{ccccc}
        \toprule
        MLP & LSTM-CNN & XGBoost & Finetuned wav2vec 2.0 & SALR \\
        \midrule
        27.44 $\pm$ 3.83 & 29.95 $\pm$ 4.15 & 37.75 $\pm$ 3.20 & 52.39 $\pm$ 2.13 & \textbf{59.23} $\pm$ 1.54 \\
        \bottomrule
    \end{tabular}
\end{table}

\subsection*{Interpretation of the Latent Space Analysis}
\begin{figure}[H]
    \centering
    \makebox[\textwidth][c]{%
        \begin{minipage}{0.9\textwidth}
            \includegraphics[width=\textwidth]{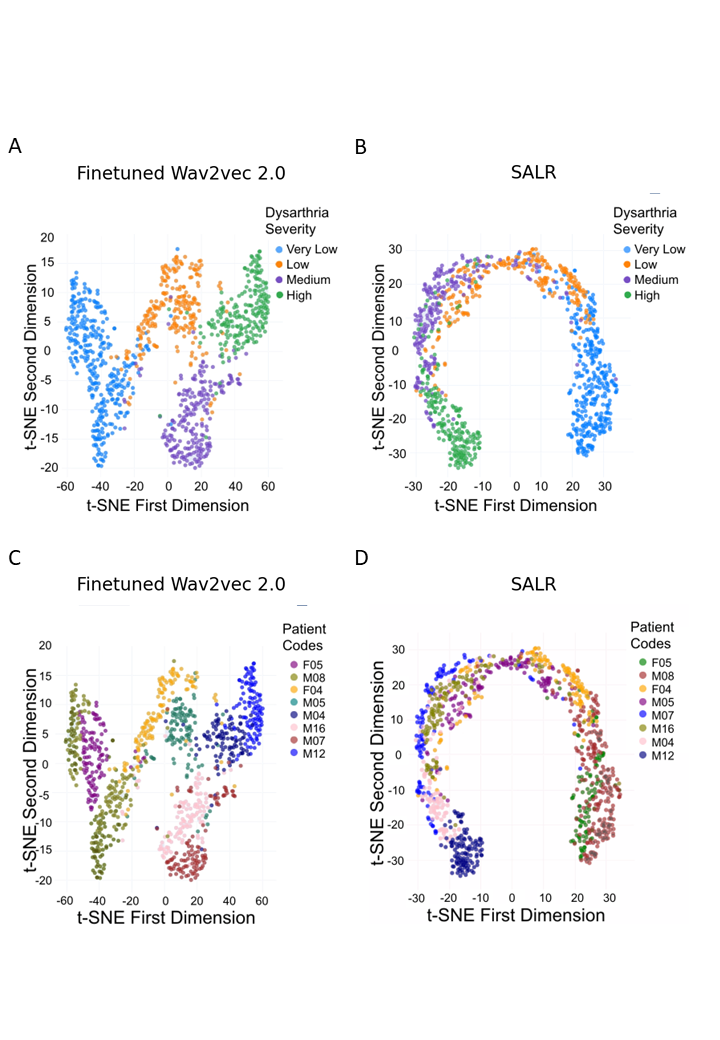}
            \caption{Visualisation of t-SNE embeddings from different models with data points coloured according to patient severity (first row) and patient code (second row). \textbf{A, C} depict the fine-tuned Wav2Vec 2.0 model, \textbf{B, D} showcase the SALR framework. These visualisations support our hypothesis that the SALR framework organises the latent space in alignment with severity levels (as seen in the first row) and disperses speaker clusters (as seen in the second row). }
            \label{fig:tsne}
        \end{minipage}
    }
\end{figure}

To further assess the impact of our frameworks on the model's representation of speech data, we conducted a t-SNE analysis on the latent space. Given t-SNE’s sensitivity
to hyperparameters, we ensured consistency by setting a perplexity of 30, a learning
rate of 30, and using 1000 iterations, as per sklearn’s t-SNE implementation. Fig \ref{fig:tsne} provides visual insights into how the models organise data with respect to dysarthria severity and speaker identity. We compare both the fine-tuned base wav2vec 2.0 model and SALR model.

For the fine-tuned base wav2vec 2.0, the latent space is unstructured with respect to the ordinal arrangement of severity levels. High-severity samples often project closely to both mid and low-severity clusters, as visualised in Fig \ref{fig:tsne}A. Moreover, there are distinct speaker clusters, as evident in Fig \ref{fig:tsne}C. Conversely, the SALR multi-task framework helps to bring an ordinal structure to the latent space as shown in Fig \ref{fig:tsne}B. Moreover, the speaker clusters appear more dispersed as shown in Fig \ref{fig:tsne}D. This indicates that our framework was successful in dispersing speaker clusters and thereby preventing the model from relying on speaker-specific cues. This enhances the model's ability to generalise across speakers while improving its discrimination of dysarthria severity levels. These observations indicate the successful achievement of our objective with the regularising contrastive loss, which aimed to disentangle speaker-specific cues from the severity classification in the latent space.

\section*{Discussion}

The primary focus of the study was developing an automated, machine learning-based tool for classifying dysarthria severity levels in a speaker-independent manner.  We fine-tuned a base wav2vec 2.0 model, a state-of-the-art self-supervised transformer model, trained on healthy speech for the task of dysarthria severity assessment. The fine-tuned  model outperformed other traditional baselines models based on XGBoost, MLP and LSTM-CNN. Further analysis suggested that the model can achieve more accurate results by avoiding any potential interference from individual idiosyncrasies unrelated to the actual severity of the condition by making the model focus more on dysarthria-specific speech features. We introduce the novel SALR multi-task framework to counteract the model’s tendency to overfit to individual speakers and improve generalisation. This augmented the fine-tuned base model performance notably, registering an accuracy of 70.48\% and an F1 score of 59.23\%. These results signify a 5.67\% boost in accuracy and a 6.84\% uptick in F1 score compared to the baseline wav2vec 2.0 model.

Deep analysis confirms that the SALR multi-task objectives does more than just improve numerical performance: it also help organise the latent space in a way that aligns with severity levels and reduce the model's reliance on speaker-specific cues which results in a subsequent boost in performance, confirming our hypothesis about the effects of the multi-task framework. Using confusion matrices for speaker-independent evaluation, we found that while the framework excels in categorising extreme severity classes, it faces challenges in distinguishing between "Low" and "Mid" severity levels. 

\subsection*{Comparison with previous works}
\begin{figure}[H]
    \centering
    \includegraphics[width=0.9\textwidth]{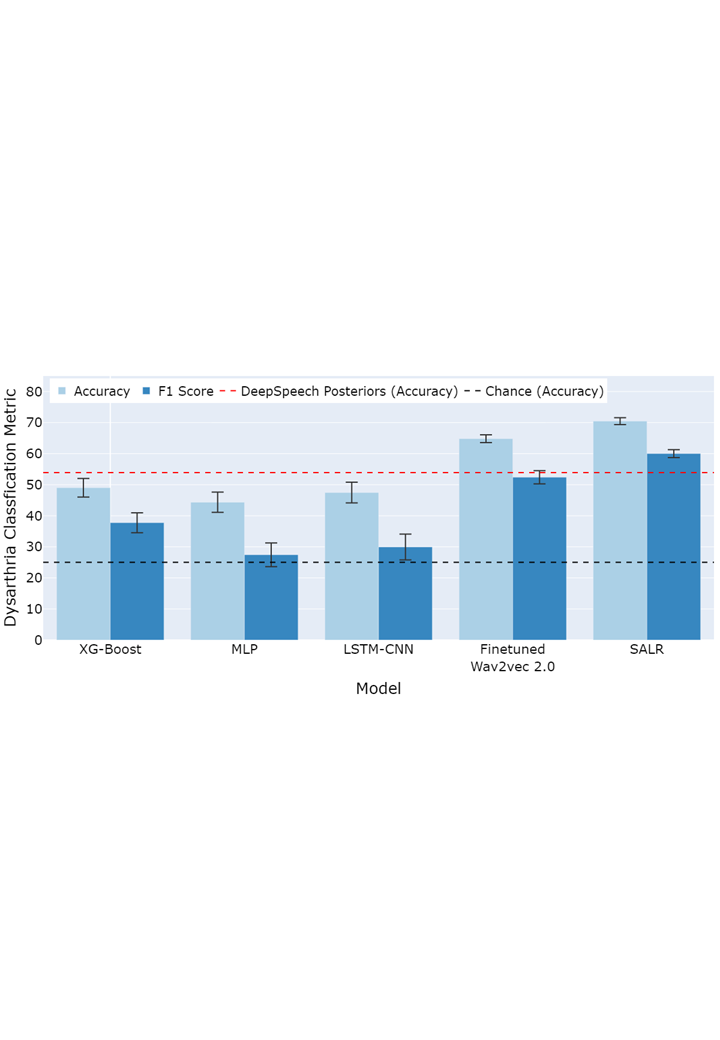}
    \caption{Comparative performance in speaker-independent multi-class dysarthria severity classification on the UA-Speech dataset(uncommon words). The bar chart illustrates the performance of our models in comparison to the existing benchmark. \cite{tripathi2020improved}.}
    \label{fig:barchart}
\end{figure}

Many existing studies have focused on speaker-dependent models where the patient in the test set was also included in the training dataset, although testing was done on words not included in the training set. However, a speaker-dependent model is of limited use in clinical settings where clinicians need to assess patients they haven't encountered before. Focusing on the task of speaker-independent multi-class classification, our innovative frameworks demonstrated a remarkable performance improvement over the prior leading work by Tripathi et al. \cite{tripathi2020improved}. The aforementioned study by Tripathi et al. utilised a combination of DeepSpeech features and Support Vector Machines (SVM) for classification, culminating in an accuracy rate of 53.90\% for a leave-one-subject-out cross-validation under speaker-independent conditions. By employing the same test set and leave-one-out subject cross-validation scheme, we ensure a fair comparison to this work and position our work appropriately within the literature. Our base fine-tuned wave2vec 2.0 model yielded a substantial enhancement in accuracy, reaching 64.81\%. Furthermore, our framework, SALR, achieved impressive accuracy rates of 70.48\% as depicted in Fig. \ref{fig:barchart}. We attribute these enhancements to the innovative integration of the SALR framework, which collectively contribute to the robustness and generalisability of our models.


\subsection*{Limitations and future work}
This study, while offering valuable insights, has its limitations, most notably in the dataset used for model training. There was a lack of patient data in the "Low" and "Mid" severity categories, resulting in an imbalanced dataset that hindered the model’s ability to accurately distinguish between mid-level severity categories. This underlines the crucial need for more comprehensive, diverse, and balanced datasets in future studies. Furthermore, our investigation did not extend to multi-lingual datasets. Given that dysarthria characteristics can vary significantly across different languages \cite{yeo2022cross}, this presents a promising avenue for future research. Exploring how dysarthria manifests in various linguistic contexts could provide deeper insights and enhance the model’s applicability across diverse populations.

The use of transformer-based frameworks, while effective, introduces challenges related to interpretability. Although our latent space visualisation provides some insights into the model’s functionality, it is imperative to adopt additional methods, such as attention heatmaps, to gain a fuller understanding of the model’s decision-making processes. This is particularly important as we progress toward automated dysarthria severity assessments, where transparency and interpretability are key. Lastly, a prospective direction for advancing the field might lie in the exploration of self-supervised pre-training tasks on dysarthric samples rather than on normal speech. However, researchers should be cognisant of the substantial computational intensity required for this approach, as exemplified by the original training of the wav2vec 2.0 model which leveraged 64 GPUs \cite{wolf2019huggingface}.

\subsection*{Clinical and research implications}
Our results underscore the potential of our multi-task frameworks in advancing the state-of-the-art in speaker-independent multi-class dysarthria severity classification, setting a new benchmark for future research and applications in this domain. The frameworks need to be further evaluated under clinical conditions with a diverse patient population to reach the ultimate goal of automated dysarthria severity assessments and improved clinical management and support for dysarthric individuals.

\section*{Conclusion}
In summary, our research addresses long-standing challenges in automated dysarthria severity assessment by introducing a multi-task framework based on the wav2vec 2.0 transformer model. This approach has led to setting a new standard for speaker-independent multi-class dysarthria severity classification evident through improved accuracy and F1 scores. With further evaluations on diverse datasets to ensure their generalisability, this research offers the potential for more accurate, efficient, and clinically relevant automated dysarthria severity assessment.

\section*{Acknowledgments}
AAF acknowledges his support by the UKRI Turing AI Fellowship (EP/V025449/1).

\section*{Authors' Contributions}
AAF conceptualised the study. LS did the machine learning analysis with technical inputs from BK \& AAF and clinical inputs from SW. LS and BK wrote the initial draft of the manuscript. All authors reviewed and edited the manuscript.

\section*{Conflicts of Interest}
None declared.


%
%
%

\newpage

\newpage

\section*{Supporting information}

\paragraph*{S1 Fig.}
\label{S1_Fig}
Comparative performance in speaker-independent multi-class dysarthria severity classification on the UA-Speech dataset using all words. 
\begin{figure}[H]
    \includegraphics[width=0.9\textwidth]{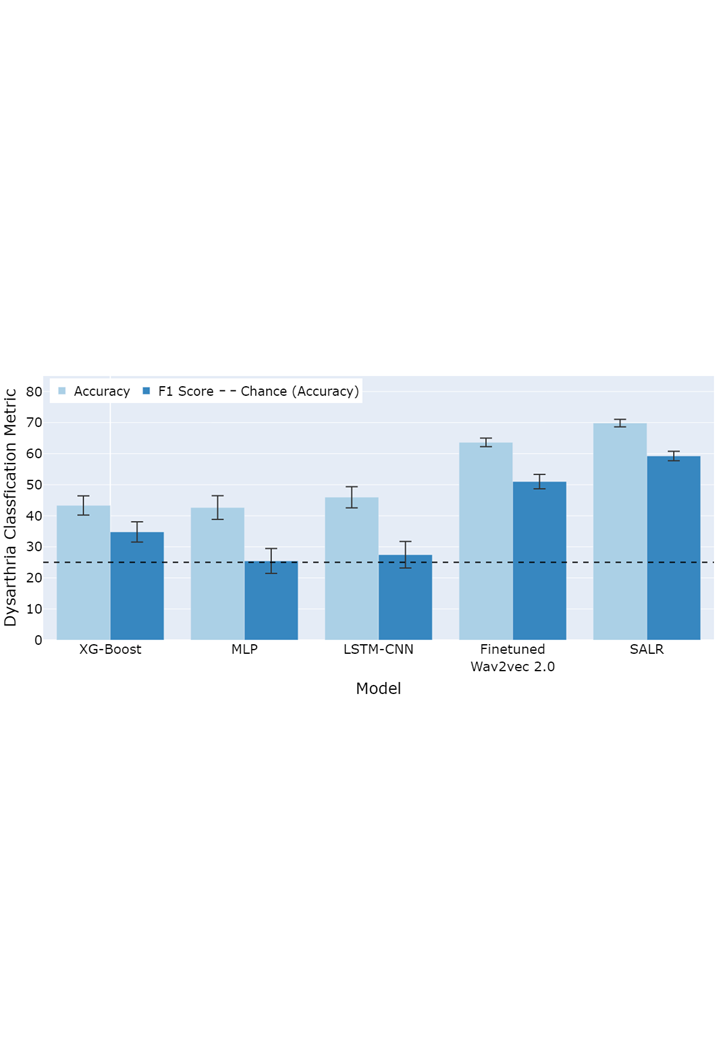}
\end{figure}

\paragraph*{S2 Fig.}
\label{S2_Fig}
Normalised confusion matrices showing the predictions on all words by \textbf{A.} fine-tuned Wav2Vec 2.0 model, \textbf{B.} SALR framework
\begin{figure}[H]
    \includegraphics[width=0.9\textwidth]{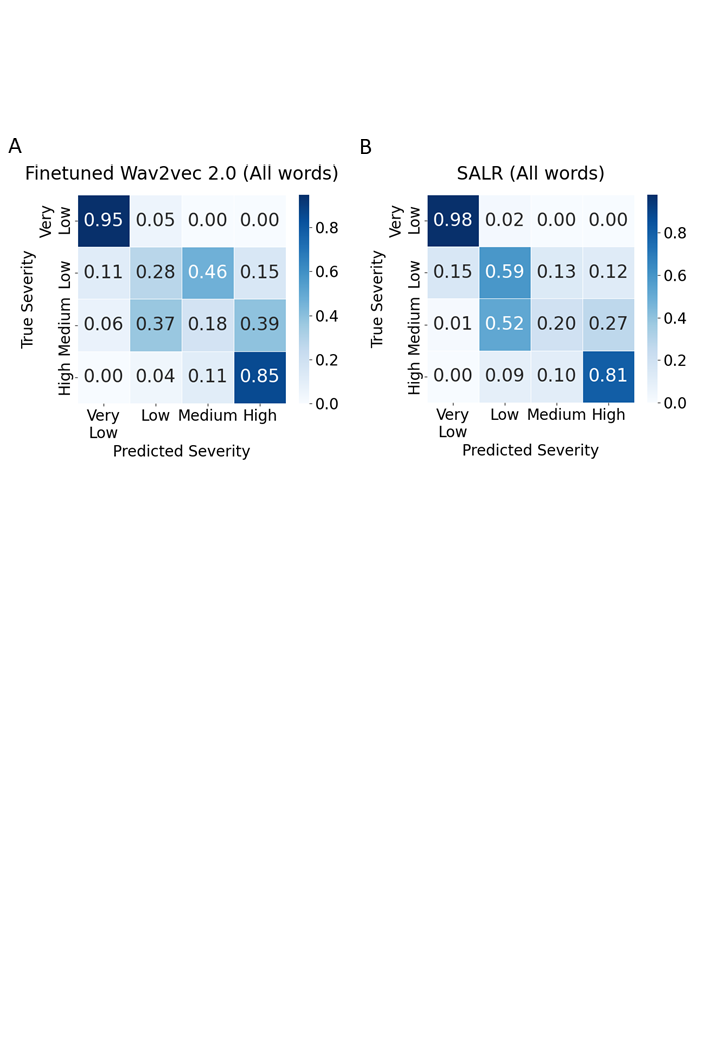}
\end{figure}






\end{document}